\documentclass{article}

\usepackage{siunitx}

\usepackage[colorinlistoftodos]{todonotes}
\usepackage{xcolor}
\usepackage{soul}

\usepackage[preprint, nonatbib]{nips_2018}

\usepackage[utf8]{inputenc} 
\usepackage[T1]{fontenc}    
\usepackage{hyperref}       
\usepackage{url}            
\usepackage{booktabs}       
\usepackage{amsfonts}       
\usepackage{nicefrac}       
\usepackage{microtype}      
\usepackage{amsmath}

\title{Technical Considerations for Semantic Segmentation in MRI using Convolutional Neural Networks}

\author{
  Arjun D.~Desai\\
  Department of Radiology\\
  Stanford University\\
  \texttt{arjundd@stanford.edu} \\
  \And
  Garry E.~Gold\\
  Departments of Radiology, Bioengineering, and Orthopedic Surgery\\
  Stanford University\\
  \texttt{gold@stanford.edu}\\
  \And
  Brian A.~Hargreaves\\
  Departments of Radiology, Electrical Engineering, and Bioengineering\\
  Stanford University\\
  \texttt{bah@stanford.edu}\\
  \And
  Akshay S.~Chaudhari\\
  Department of Radiology\\
  Stanford University\\
  \texttt{akshaysc@stanford.edu}\\
}

\begin{document}

\maketitle

\begin{abstract}
High-fidelity semantic segmentation of magnetic resonance volumes is critical for estimating tissue morphometry and relaxation parameters in both clinical and research applications. While manual segmentation is accepted as the gold-standard, recent advances in deep learning and convolutional neural networks (CNNs) have shown promise for efficient automatic segmentation of soft tissues. However, due to the stochastic nature of deep learning and the multitude of hyperparameters in training networks, predicting network behavior is challenging. In this paper, we quantify the impact of three factors associated with CNN segmentation performance: network architecture, training loss functions, and training data characteristics. We evaluate the impact of these variations on the segmentation of femoral cartilage and propose potential modifications to CNN architectures and training protocols to train these models with confidence.
\end{abstract}

\section{Introduction}
Magnetic resonance imaging (MRI) provides high spatial resolution and exquisite soft tissue contrast, leading to its pervasive use for visualization of tissue anatomy. Using MR images for quantitative analysis of tissue-specific information is critical for numerous diagnostic and prognostic protocols. The gold-standard for high fidelity region annotation is manual tissue segmentation, which can be time-consuming and prone to inter-reader variations \cite{hoyte2011segmentations, bogner2012readout}. Thus, there has always been great interest in developing fully-automated tissue segmentation techniques that are robust to small image variations \cite{dam2015automatic, bauer2013survey}.

A common application of segmentation is the segmentation of articular cartilage for studying changes associated with onset and development of osteoarthritis (OA) \cite{erhart2015new, dunn2004t2}. Recent advances in MRI have focused on developing non-invasive morphological and compositional biomarkers for tracking the onset and progression of OA. There is promising evidence suggesting that changes in cartilage morphology and composition may serve as early-OA biomarkers \cite{eckstein2013quantitative, welsch2008cartilage}. Despite its potential, accurate measurement of cartilage morphology entails tedious manual segmentation of the fine structure of cartilage in hundreds of MRI slices and patients \cite{Eckstein_2006}. Though automatic segmentation of femoral cartilage is of great interest, the tissue's thin morphology and low contrast with surrounding tissue structures makes automatic segmentation challenging. 

Traditional automatic segmentation approaches have utilized 3D statistical shape modeling or multi-atlas segmentations modulated by anisotropic regularizers \cite{seim2010model, shan2014automatic}. However, such techniques are highly sensitive to deformations in knee shape, which can be caused by variations in patient knee size and in incidence and progression of pathology \cite{pedoia2016segmentation}. Advances in deep learning and convolution neural networks (CNNs) have shown great potential for enhancing the accuracy of cartilage segmentation \cite{liu2018deep, norman2018use}. However, due to the stochastic nature of deep learning and the multitude of training parameters (hyperparameters) that can be fine-tuned for any given problem, developing analytic estimations of network behavior is challenging \cite{lecun2015deep, shwartz2017opening}. As a result, practical design choices for optimizing CNN performance for segmentation in MRI, especially for femoral cartilage segmentation, have been under-explored.

Often, CNN architectures are modified in the hope of increasing overall accuracy and generalizability while minimizing inference time. In the case of the popular ImageNet challenge \cite{russakovsky2015imagenet} for natural image classification, classification accuracy and generalizability have varied considerably with changes in network architecture \cite{szegedy2015going, he2016deep}. Additionally, while 2D architectures have been effective at slice-by-slice segmentation of medical images, recent works have also utilized volumetric architectures, which take 3D volumes as inputs to potentially add through-plane (depth) contextual cues to improve segmentation \cite{kamnitsas2017efficient, cciccek20163d}. However, the extent to which network structure and input depth impact semantic segmentation in medical imaging remains unclear.

Variations in CNN training protocol may also affect network performance. As network weights are optimized with respect to the gradient of the training loss function, the selection of loss function may dictate network accuracy. In particular for segmentation, where foreground-background class imbalance is common, loss functions, such as weighted cross-entropy and soft Dice loss, are often chosen to minimize the impact of class imbalance \cite{tian2017deep, baumgartner2017exploration}. In addition, supervised CNN training requires both large training sets and corresponding high-fidelity segmentation masks, which are difficult to produce. In cases of limited training data, data augmentation is a common practice for artificially increasing variability of training data to reduce overfitting and promote generalizability \cite{perez2017effectiveness, krizhevsky2012imagenet}. Moreover, MRI volumes can be acquired with varying fields of view (FOVs), resulting in different matrix sizes. A commonly reported advantage of fully convolutional network (FCN) CNN architectures is their ability to infer on images or volumes of arbitrary matrix sizes not specifically utilized during the training process \cite{long2015fully}.

In this study, we investigate three factors associated with the performance and generalizability of segmentation CNNs: network architecture, training loss functions, and data extent. While performance is quantified by traditional segmentation accuracy metrics, we also quantify the generalizability of a network using the sensitivity of applying trained networks to segment MR images at varying FOVs. All experiments were conducted on segmentation of femoral cartilage, a challenging localization problem and a relevant target for studying OA. We seek to quantify performance variations induced by these three factors to motivate how CNN segmentation models can be built, trained, and deployed with confidence.

\section{Methods}

\subsection{Dataset}
\label{sec:methods_dataset}
Data for this study were acquired from the Osteoarthritis Initiative (OAI) (http://oai.epi-ucsf.org), a longitudinal study for studying osteoarthritis progression \cite{Peterfy_2008}. 3D sagittal double echo steady state (DESS) datasets along with their corresponding femoral cartilage segmented masks were utilized in this study (relevant scan parameters: FOV=14cm, Matrix=384$\times$307 (zero-filled to 384$\times$384), TE/TR=5/16ms, 160 slices with a thickness of 0.7mm) \cite{Peterfy_2008}. This dataset consisted of 88 patients with Kellgren-Lawrence (KL) OA grades between 1 and 4 \cite{kellgren1958osteo}, measured at two time points (baseline and 1 year) for a total of 176 segmented volumes. These patients were randomly split into cohorts of 60 patients for training, 14 for validation, and 14 for testing, resulting in 120, 28, and 28 volumes used during training, validation, and testing, respectively. An approximately equal distribution of KL grades was maintained among all three groups (Supporting Table S1). 

\subsection{Data Pre-processing}
\label{sec:methods_data_preprocessing}
All DESS volumes in \S\ref{sec:methods_dataset} were downsampled by a factor of 2 in the slice dimension (to dimensions of 384$\times$384$\times$80) prior to network training and inference to increase SNR and reduce computational complexity, justified by previous studies reporting that approximately 1.5mm slices are adequate for cartilage morphometry \cite{Eckstein_2006}. Images were downsampled using sum-of-squares combinations, and the corresponding masks were downsampled by taking the union of the masks to compensate for partial volume artifacts. The volume was then cropped in-plane to 288$\times$288 by calculating the center of mass (COM) and centering the cropped region 50 pixels in the superior (up) direction and 20 pixels in the anterior (left) direction to bias the COM to femoral cartilage and away from the tibia and posterior muscles. The volume was then cropped to remove 4 slices from both the medial and lateral ends, resulting in volumes of dimensions 288$\times$288$\times$72. All scans were subsequently volumetrically zero-mean whitened ($\mu$=0, $\sigma$=1) by subtracting the image volume mean and scaling by the image volume standard deviation.

All data pre-processing was conducted using MATLAB (MathWorks, Natick, MA). These training, validation, and testing sets were used for all experiments unless otherwise indicated.

\begin{figure}[bt]
\centering
\includegraphics[width=10cm]{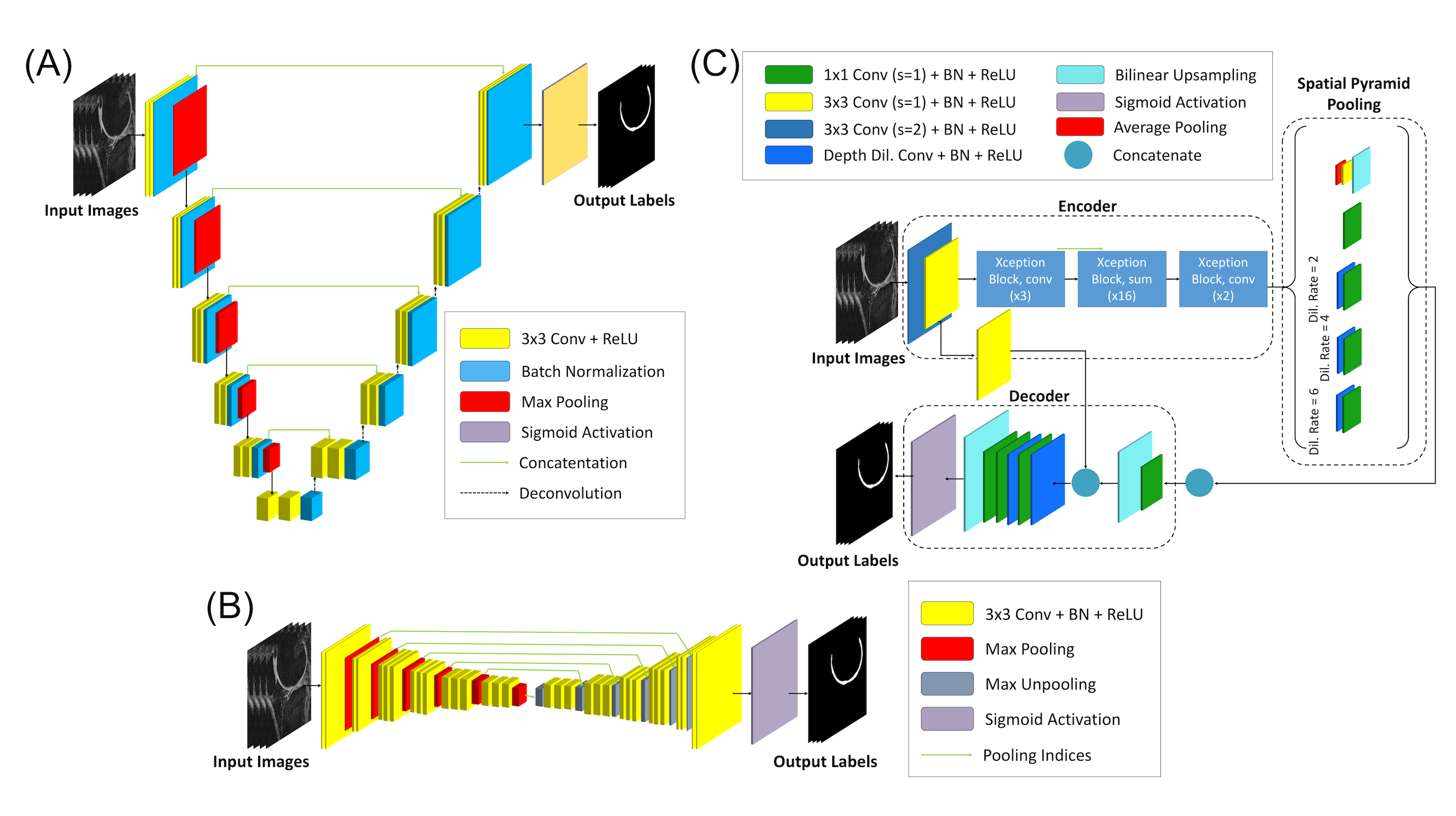}
\caption{Three encoder-decoder fully convolutional network architectures adopted for femoral cartilage segmentation. U-Net (A) uses skip connections to concatenate weights from the encoder to the decoder while SegNet (B) passes pooling indices to the decoder to reduce computational complexities of weight concatenation. Unlike traditional encoder-decoder architectures, DeeplabV3+ (C) uses spatial pyramid pooling and atrous (dilated) convolutions to extract latent feature vectors at multiple fields of view.}
\label{fig:architectures}
\end{figure}

\subsection{Network Architecture}
\label{sec:methods_network_architecture}
In this experiment, we wanted to evaluate the sensitivity of the semantic segmentation task to different, popular CNN architectures. We selected three general 2D FCN architectures for analysis: U-Net, SegNet, and DeeplabV3+ \cite{ronneberger2015u, badrinarayanan2015segnet, chen2018deeplab}. These FCN architectures utilize variations of the encoder-decoder model for semantic segmentation for extracting features at different spatial fields of view.

The U-Net architecture implements an encoder-decoder model using max-pooling and transpose convolutions to downsample and upsample feature maps (Figure ~\ref{fig:architectures}a). In this structure, the number of network filters increases exponentially as a function of network depth. The U-Net also relies on deep skip connections by concatenating encoder outputs to the decoding layers in order to share spatial cues between the two and to propagate the loss efficiently at different network depths \cite{glorot2010understanding, pascanu2013difficulty}. SegNet uses a similar encoder-decoder architecture but passes pooling indices to upsampling layers to avoid the overhead of copying encoder weights (Figure ~\ref{fig:architectures}b). In contrast to using max-pooling to promote spatial invariance and to downsample feature maps, DeeplabV3+ implements `Xception` blocks \cite{chollet2017xception} and spatial pyramid pooling with dilated convolutions to capture a larger receptive field without increasing the parameter size (Figure ~\ref{fig:architectures}c). Instead of transposed convolutions, the DeeplabV3+ decoder uses bilinear upsampling to upsample the features to the input image size. While the U-Net and SegNet have shown promise for musculoskeletal MRI semantic segmentation \cite{liu2018deep, norman2018use}, DeeplabV3+ has been primarily utilized for natural image segmentation \cite{chen2017rethinking} and has seen limited use in segmentation of medical images. 

As a baseline comparison, all architectures were trained for 100 epochs and subsequently fine-tuned for 100 epochs following training hyperparameters detailed in Table ~\ref{table:default_hyperparameters}.

\begin{table}[bt]
\caption{Default hyperparameters used for network training.}
\label{table:default_hyperparameters}
\begin{tabular}{lccccc}
\toprule
Architectures    & BS    & initial LR    & LR step-decay (DF, DP)    & Optimizer     &    Initialization \\
\midrule
U-Net & 35 & 1e-2 & 0.8, 1 & Adam & He\\
SegNet & 15 & 1e-3 & N/A & Adam & He\\
DeeplabV3+ & 12 & 1e-4 & N/A & Adam & He\\
\bottomrule  
\end{tabular}\par
BS, mini-batch size; LR, learning rate; DF, drop factor; DP, drop period (epochs)
\end{table}

\subsection{Volumetric Architectures}
\label{sec:methods_volumetric_architectures}
In this experiment, we trained a 2.5D \cite{roth2014new} and 3D U-Net architecture for femoral cartilage segmentation. The 2.5D network uses a stack of $t$ continuous slices in a scan to generate a segmentation mask for the central slice (additional details are described in the supplemental information). Three 2.5D networks with inputs of thickness $t$=3,5,7 were trained. 

In contrast, a 3D network outputs a segmentation on all $t$ slices. As all operations are applied in 3D, the 2$\times$ max-pooling applied in the through-plane direction constrains the input to have $t=2^{N_p}$ slices, where $N_p$ refers to the number of pooling steps. To maintain an identical number of pooling steps as the 2D and 2.5D networks ($N_p=5$), the 3D U-Net was trained using 32 slices of the volume as an input ($t=32$). As a result, the scans in the training dataset described in \S\ref{sec:methods_data_preprocessing} were also cropped by an additional 4 slices from the medial and lateral ends, resulting in volumes with 64 slices. Memory constraints of the hardware necessitated that this volume be further divided into two 3D subvolumes of size 288$\times$288$\times$32 and that the batch size be reduced to 1. The 2D and 2.5D networks had an exponentially increasing number of filters ranging from 32$\xrightarrow{}$1024, while the 3D network had filters ranging from 16$\xrightarrow{}$512 to accommodate for the same network depth. All networks maintained a comparable number of weights (Supporting Figure S2). 

\subsection{Loss Function}
\label{sec:methods_loss_function}
As trainable parameters in a network update with respect to the loss function, we hypothesize that a relevant loss function is critical for any learning task. Traditionally, binary cross-entropy has been used for binary classification tasks. However, class imbalance has shown to limit the optimal performance in cases of general cross-entropy losses \cite{kamnitsas2017efficient}. We selected three additional loss functions commonly used for segmentation in cases of class imbalance for comparison: soft Dice loss \cite{milletari2016v}, weighted cross-entropy (WCE), and focal loss ($\gamma$=3) \cite{lin2018focal}, as described additionally in the supplementary. 

In this experiment, four models using the U-Net architecture were trained using the four loss functions described above with the training, validation, and testing sets described in \S\ref{sec:methods_dataset}.

\begin{figure}[bt]
\centering
\includegraphics[width=10cm]{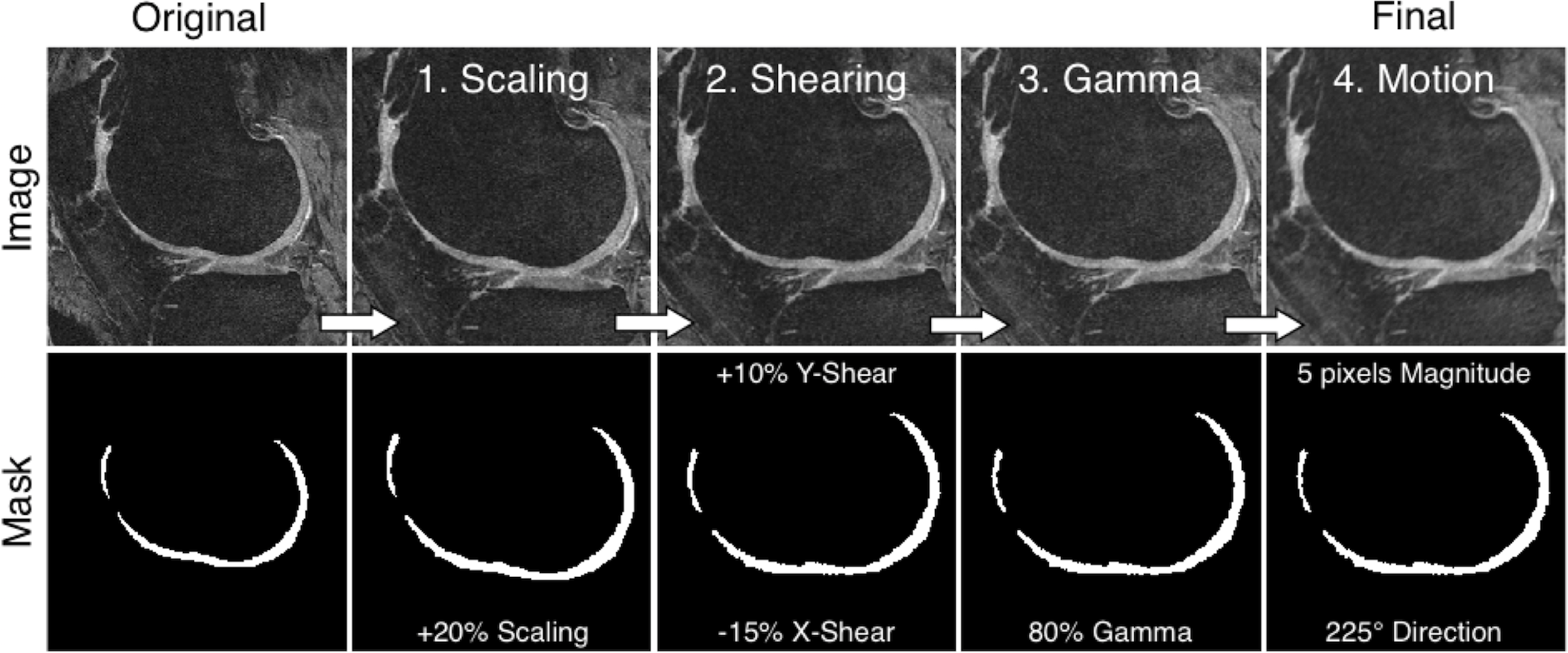}
\caption{An example augmented, final 2D slice is generated from an original 2D slice by sequentially applying four feasible transformation factors: scaling, shearing, gamma, and motion. Parameters for all four factors are sampled uniformly at random.}
\label{fig:augmentation}
\end{figure}

\subsection{Data Augmentation}
\label{sec:methods_data_augmentation}
To qualify the effect of data augmentation on model generalizability, we trained the standard U-Net architecture with and without augmented training data.

Each 2D slice and corresponding mask in the training volume were randomly augmented to add heterogeneity to the training dataset. The augmentation procedure consisted of sequential transformations of the original image and masks with: 1. zooming (between 0-10\%), 2. shearing (between -15$^{\circ}$ to 15$^{\circ}$ in both the horizontal and vertical directions), 3. gamma variations (between 0.8-1.1 for simulating varying contrasts), and 4. motion blur (between 0-5 pixels in magnitude and 0$^{\circ}$ to 360$^{\circ}$ in direction). Parameters for each transformation were chosen uniformly at random within the specified ranges, with an example slice shown in Figure ~\ref{fig:augmentation}. These specific augmentation methods and magnitudes were chosen to mimic typically encountered physiological and imaging variations. No augmentations were applied to the scans in the validation and test sets.

All 2D slices were augmented fourfold, resulting in the augmented dataset consisting 5x the data in the non-augmented training set. To overcome this discrepancy while training separate networks with and without augmented data, the networks trained using augmented data were trained for 5x shorter than those trained using non-augmented data (20 epochs total). 

\subsection{Generalizability to Multiple Fields of View}
\label{sec:methods_FOV}
In this experiment, we compare the differences in network performance on scans at different FOVs with the same underlying image resolution. In addition to the inference dataset (V0) cropped to a volume of ($288\times288\times72$) described in \S\ref{sec:methods_dataset}, three new test sets were created with different degrees of cropping: V1 ($320\times320\times80$), V2 ($352\times352\times80$), and V3 ($384\times384\times80$). 

As data augmentation is hypothesized to increase network generalizability, we compared the performances of the U-Net models trained using non-augmented and augmented data as specified in \S\ref{sec:methods_data_augmentation} among the four test sets (V0-V3).

\subsection{Training Data Extent}
\label{sec:methods_data_limitation}
Performance of CNNs has also been shown to be limited by the extent (amount) of data available for training \cite{rafiq2001neural}. To explore the relationship between the extent of training data and network accuracy, we trained each of the three base network architectures in \S\ref{sec:methods_network_architecture} on varying sized subsets of the training data. The original training set consisting of 60 patients was randomly sampled (with replacement) to create 3 additional sub-training sets of 5, 15, and 30 patients with similar distributions of KL grades (Supporting Table S3). The same validation and testing sets described in \S\ref{sec:methods_dataset} (with 14 patients, each at two time points) were used to assess the generalizability of the networks. 

The network trained on the complete sample of training data (60 patients) was trained for 100 epochs. To ensure that all sub-sampled networks maintained an equal number of backpropagation steps to update filter weights, we scaled the number of epochs by the ratio of the fully sampled patient count (60) to the number of patients in the sub-training set. As a result, networks trained on 5, 15, and 30 patients were trained for 1200, 400, and 200 epochs respectively. Experiments were repeated 3 times each (with Python seeds 1, 2, and 3) to enhance reproducibility and to minimize the stochasticity of random network weight initializations.

\subsection{Network Training Hyperparameters}
For all experiments, convolutional layers with rectified linear unit (ReLU) activations were initialized using the "He" initialization \cite{nair2010rectified, he2015delving}. Training was performed using the Adam optimizer with default parameters ($\beta_1=0.9$, $\beta_2=0.999$, $\epsilon$=1e-8) with random shuffling of mini-batches using a Tensorflow backend \cite{kingma2014adam, abadi2016tensorflow}. All neural network computations were performed on 1 Titan Xp graphical processing unit (GPU, NVIDIA, Santa Clara, CA) consisting of 3,840 CUDA cores and 12GB of GDDR5X RAM.

Due to the randomness of the training processes, we empirically determined a pseudo-optimal set of hyperparameters for training each network. To reduce large variances in training batch normalization layers caused by small mini-batch sizes, the largest mini-batch size that could be loaded on the Titan Xp GPU was used for each network. The initial learning rate and use of step decay was also empirically determined based on the network architecture. Table ~\ref{table:default_hyperparameters} details the hyperparameters used with each network architecture. Networks were trained using the soft Dice loss, unless otherwise specified.

\subsection{Quantitative Comparisons}
For each experiment, the model that resulted in the best loss on the validation dataset was used for analysis on the testing dataset. During testing, output probabilities of femoral cartilage ($p_{FC}$) were thresholded at 0.5 to create binary femoral cartilage segmentations ($p_{FC}\leq0.5 \xrightarrow{} 0$, $p_{FC}>0.5 \xrightarrow{} 1$). No additional post-processing was performed on the thresholded masks. 

Segmentation accuracy was measured on the testing dataset using three metrics - Dice similarity coefficient (DSC), volumetric overlap error (VOE), and coefficient of variation (CV) \cite{taha2015metrics}. High accuracy segmentation methods maximize DSC (a maximum of 100\%) while minimizing VOE and CV (a minimum of 0\%). The segmentation masks obtained from the OAI dataset served as ground truth. Statistical comparisons between the inference accuracy of different models were assessed using Kruskal-Wallis tests, and corresponding Dunn post-hoc tests, ($\alpha=0.05$). All statistical analyses were performed using the SciPy (v1.1.0) library \cite{jones2014scipy}.

Additionally, changes in network performance in the slice (depth-wise) direction were visualized using graphs termed depth-wise region of interest distribution (dROId) plots. The normalized depth-wise field of view (dFOV) spanning the region of interest is defined as the ordered set of continuous slices containing femoral cartilage according to ground truth manual segmentation, where, in the set, the first slice corresponds to medial side (dFOV=0\%) and the last slice corresponds to the lateral side (dFOV=100\%). All volumes were mirrored to follow this convention.

\begin{table}[bt]
\centering
\caption{A summary of network performance (mean (standard deviation)) in base and volumetric architecture, loss function, and data augmentation experiments. Models with Dice score coefficient  (DSC) accuracy and volumetric overlap error (VOE) significantly (p<0.05) than corresponding metric of \textit{all} other models in the given experiment are marked with *. Models with all metrics signficantly different (p<0.01) than corresponding metric of \textit{all} other models in the given experiment are marked with **. Best performing networks in each experiment category are bolded.}
\includegraphics[width=13cm]{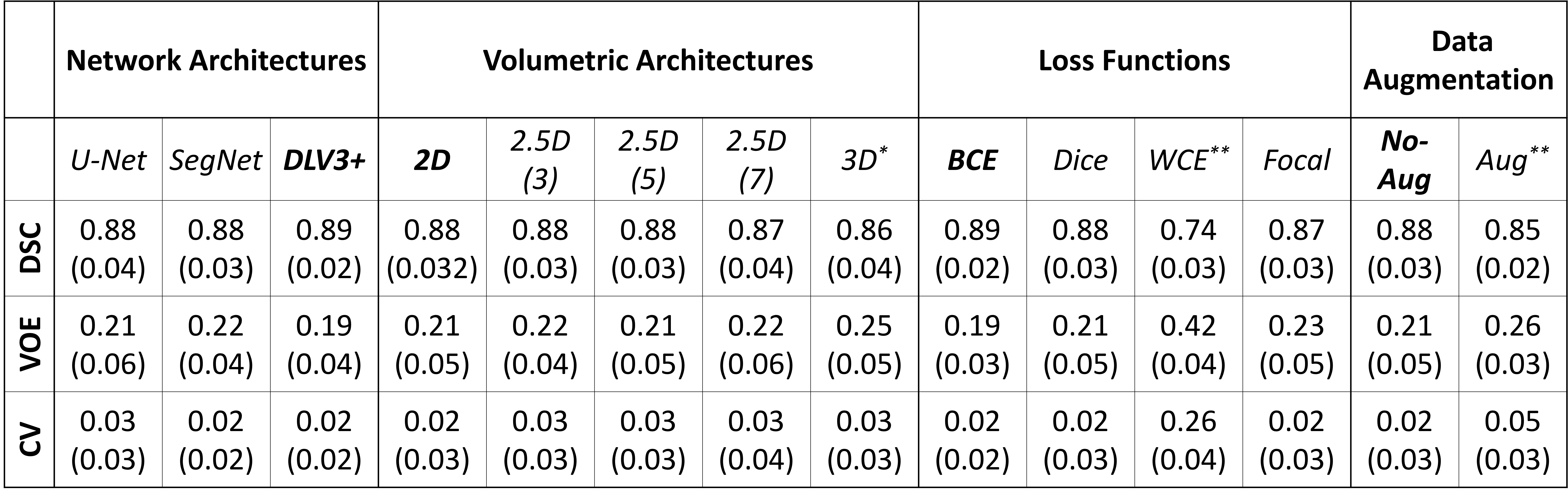}
\label{table:perf-summary}
\end{table}

\section{Results}
All performance results (except data limitation) are summarized in Table ~\ref{table:perf-summary}.

\begin{figure}[bt]
\centering
\includegraphics[width=10cm]{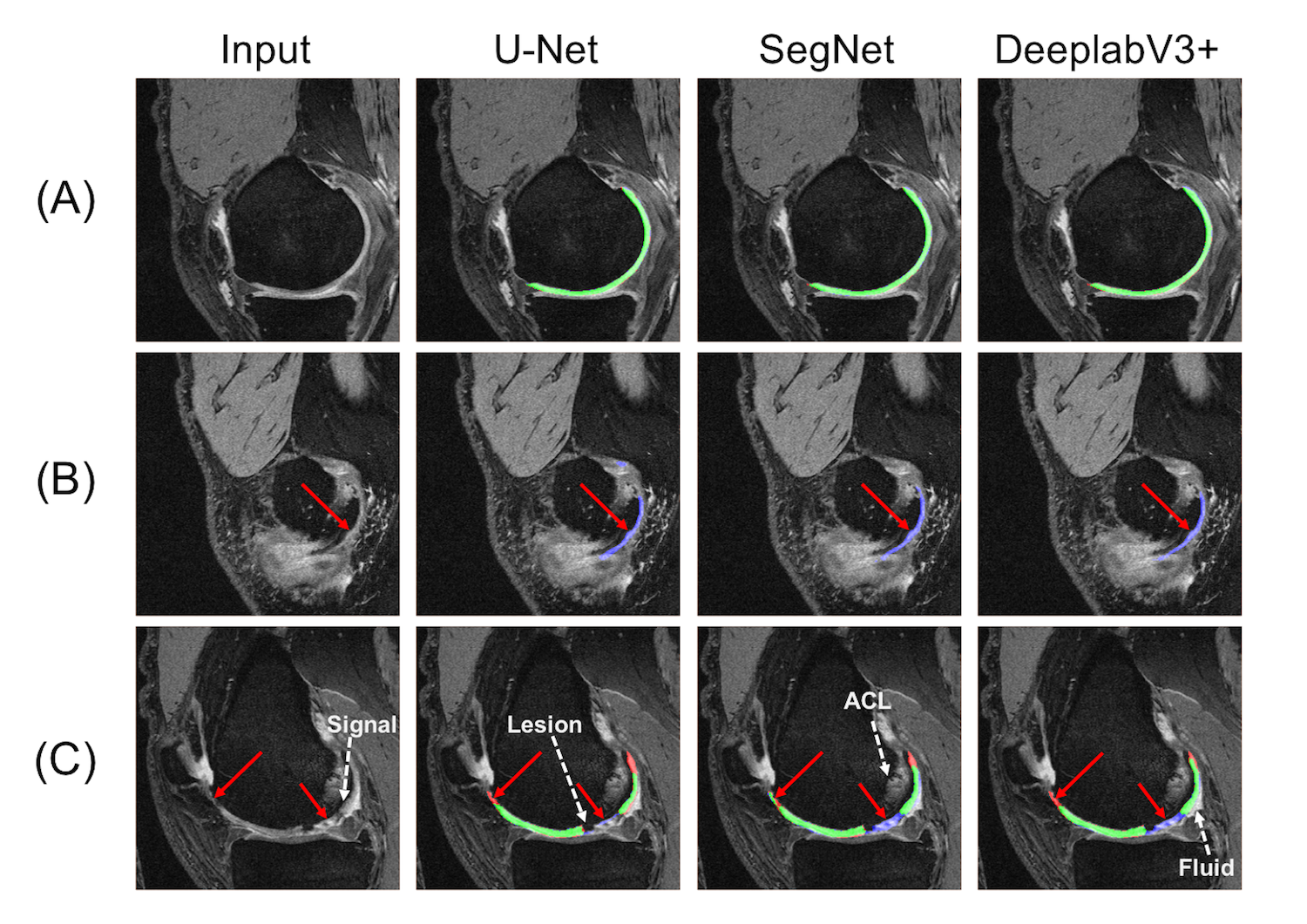}
\caption{Sample segmentations from three FCN architectures (U-Net, SegNet, DeeplabV3+) with true-positive (green), false-positive (blue), and false-negative (red)
overlays. Despite statistically significant difference between the performance of U-Net and the other two architectures, there is minimal visual variation between
network outputs. Thick, continuous cartilaginous regions (A) have considerably better performance throughout the entire region, including edge pixels. Failures (red arrows) occur in regions of thin, disjoint femoral cartilage common in edge (B) and medial-lateral transition slices (C). However, (C) shows all networks successfully handled challenging slices that include difficult to segment anatomy (white arrows), such as cartilage lesions, heterogeneous signal, and proximity to anatomy with similar signal (ACL, fluid).}
\label{fig:arch-comparison}
\end{figure}

\begin{figure}[bt]
\centering
\includegraphics[width=10cm]{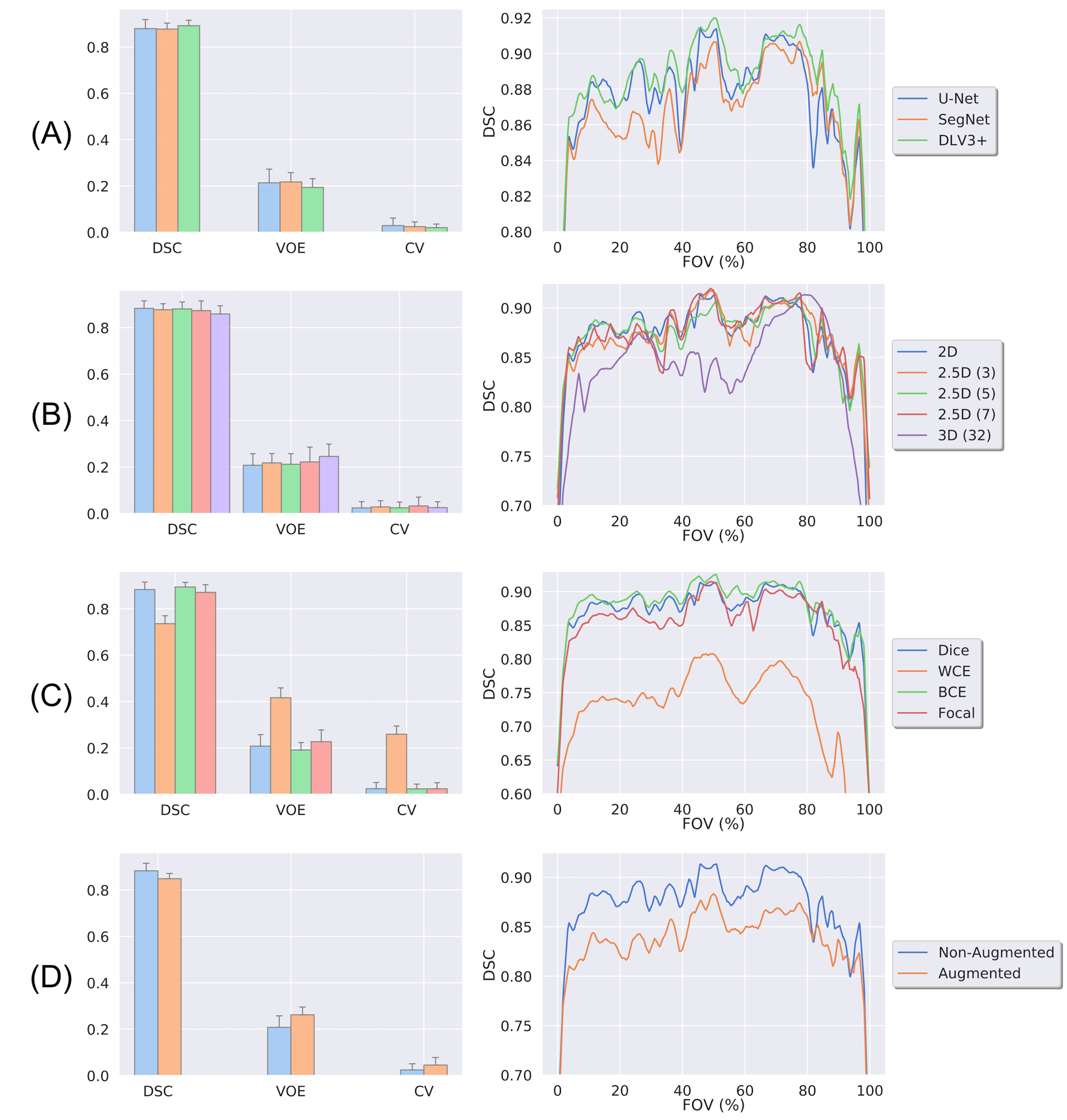}
\caption{Performance bar graphs and depth-wise region of interest distribution (dROId) plots for convolutional neural network models with different (A) network architectures, (B) volumetric architectures, (C) training loss functions, and (D) training data augmentations. The field of view defined by the region of interest in dROId plots is normalized (0-100\%) to map performance at knee-specific anatomical locations despite variations in patient knee size.}
\label{fig:ROI-graph}
\end{figure}

\subsection{Network Architecture Comparison}
A comparison of the performance of the U-Net, SegNet, and DeeplabV3+ architectures on sample slices is shown in Figure ~\ref{fig:arch-comparison}. All three base architectures maintained high fidelity in segmenting slices containing thick cartilage structures (Figure ~\ref{fig:arch-comparison}A). However, all networks had worse performance in slices containing regions of full-thickness cartilage loss and denuded subchondral bone, edge slices, and medial-lateral transition regions (Figure ~\ref{fig:arch-comparison}B,C). Despite lower accuracy in these regions, these networks accurately segmented slices with heterogeneous signal caused by pathology and proximity to anatomy with similar signal (Figure ~\ref{fig:arch-comparison}C). Performance decreased at edge regions (dFOV\textasciitilde[0, 10]\%, [90, 100]\%) and at the medial-lateral transition region (dFOV\textasciitilde[55, 65]\%) as seen in the dROId plot in Figure ~\ref{fig:ROI-graph}A. There was no significant difference in the performance of U-Net, SegNet, and DeeplabV3+ models as measured by DSC (p=0.08), VOE (p=0.08), and CV (p=0.81).

\subsection{Volumetric Architectures Comparison}
Results of the 2D, 2.5D, and 3D U-Net architectures showed no significant difference between the performance of 2D U-Net and that of the three versions ($t$=3,5,7) of the 2.5D U-Net. The 2D U-Net, however, did perform significantly better than the 3D U-Net (DSC,VOE-p<0.05). There were also no significant differences (p=1.0) in the performance between the 2.5D architectures using inputs of different depths ($t$=3,5,7). Decreased DSC at edge and medial-lateral transition regions was indicative for all models as seen on the dROId plot (Figure ~\ref{fig:ROI-graph}B). In the 3D U-Net model, DSC was greater in the lateral compartment of the knee (dFOV\textasciitilde[60,90]\%) compared to that of the medial compartment (dFOV\textasciitilde[15, 45]\%). Among 2D and 2.5D networks, performance in the lateral and medial regions was comparable. 

\begin{figure}[bt]
\centering
\includegraphics[width=7cm]{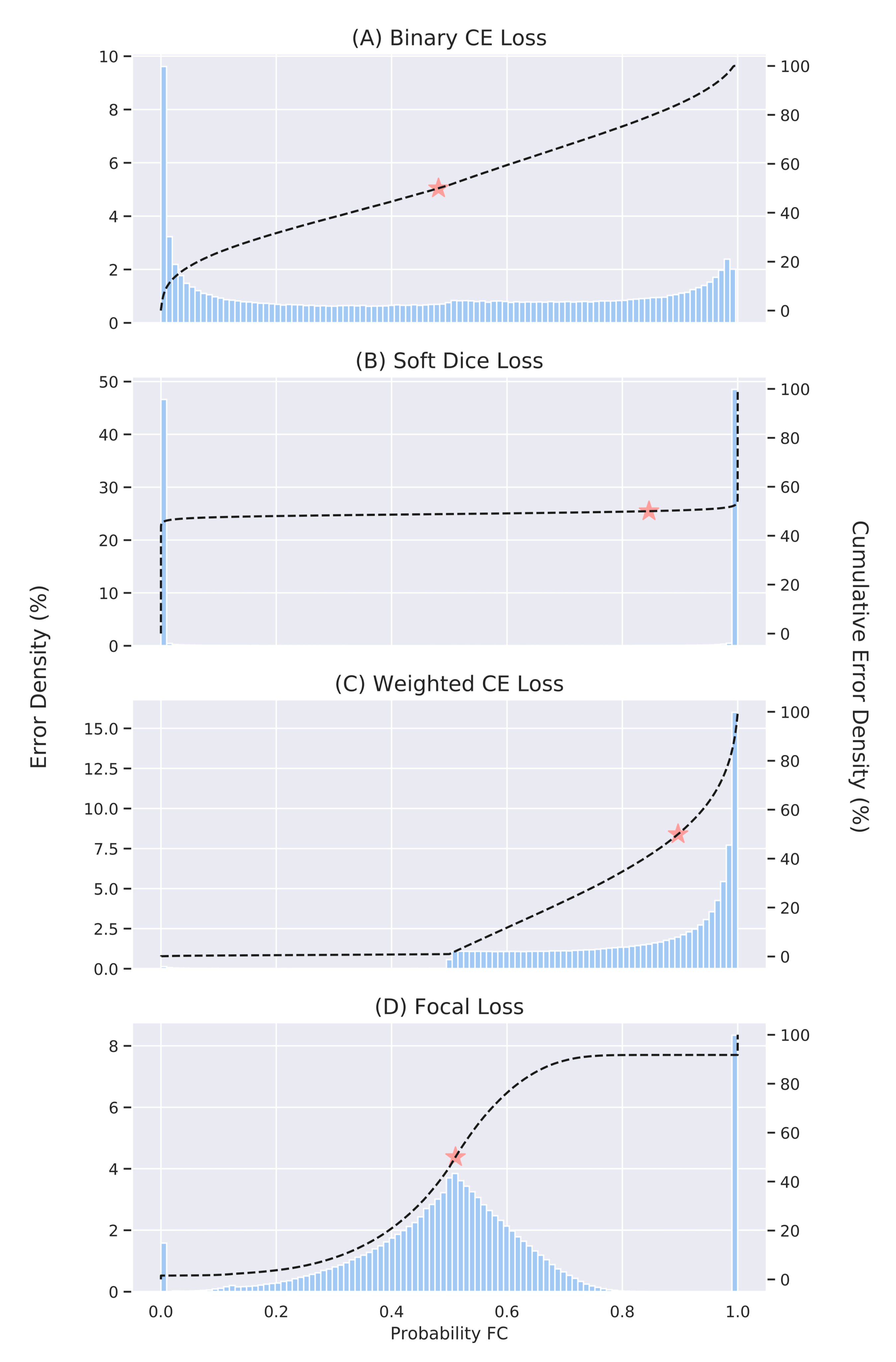}
\caption{A summary of performances of networks trained on (A) binary cross-entropy (BCE), (B) soft Dice, (C) weighted cross-entropy (WCE), and (D) focal losses. The estimated median probability ($p^*_{FC}$, defined as $CDF(p^*_{FC})=0.5$, is marked by the red star. BCE has peak errors at $p_{FC}$=0, $p_{FC}$=1, with relatively uniform number of errors at low confidence probabilities ($0.25 \leq p_{FC} \leq 0.75$). Soft Dice loss has a clear bi-modal distribution with peaks at $p_{FC}$=0,1, with negligible errors at low confidence probabilities. Almost all WCE errors were false-positives ($p_{FC}$>0.5), with increasing error density at higher probabilities. While focal loss exhibits similar peaks at $p_{FC}$=0,1, as soft Dice and BCE, ~90\% of the error density is centered around low confidence probabilities.}
\label{fig:loss-distribution}
\end{figure}

\subsection{Loss Function Comparison}
Performance differences between BCE, soft Dice, and focal losses were negligible, but all three losses significantly outperformed WCE (p<5e-10) across all slices (Figure  ~\ref{fig:ROI-graph}C).

Using the WCE loss model for inference, the incidence rate of false-positives (misclassifying a background pixel as femoral cartilage) was significantly greater (p<2e-10) than the incidence rate of false-negatives (misclassifying a femoral cartilage pixel as background). Over 99\% of the WCE model errors were false-positives (Figure ~\ref{fig:loss-distribution}C). The pixel-wise error distribution, as measured on the test set (V0) appeared correlated to the output probability of femoral cartilage ($0\leq p_{FC}\leq1$), which may be an indicator of network confidence in classifying a pixel as femoral cartilage.

For BCE, soft Dice, and focal losses, the difference between the false-positive and false-negative rates were not significant (p>0.4). The incidence of errors is also symmetrically distributed around the threshold probability with medians $p^*_{FC}$ of 0.48, 0.84, and 0.51, respectively (Figure ~\ref{fig:loss-distribution} A,B,D). The error rate in BCE was relatively uniform across all probabilities while the distribution of error rates in soft Dice loss is primarily bi-modal with peaks at $p_{FC}=0$ and $p_{FC}=1$. The focal loss error distribution was more densely centered around $p_{FC}=0.5$.

\subsection{Data Augmentation Comparison}
The use of augmented training data significantly decreased network performance (p<0.001) compared to the augmented training data set (Figure ~\ref{fig:ROI-graph}B). The performance was also consistently lower at other regions of the knee.

\begin{figure}[bt]
\centering
\includegraphics[width=10cm]{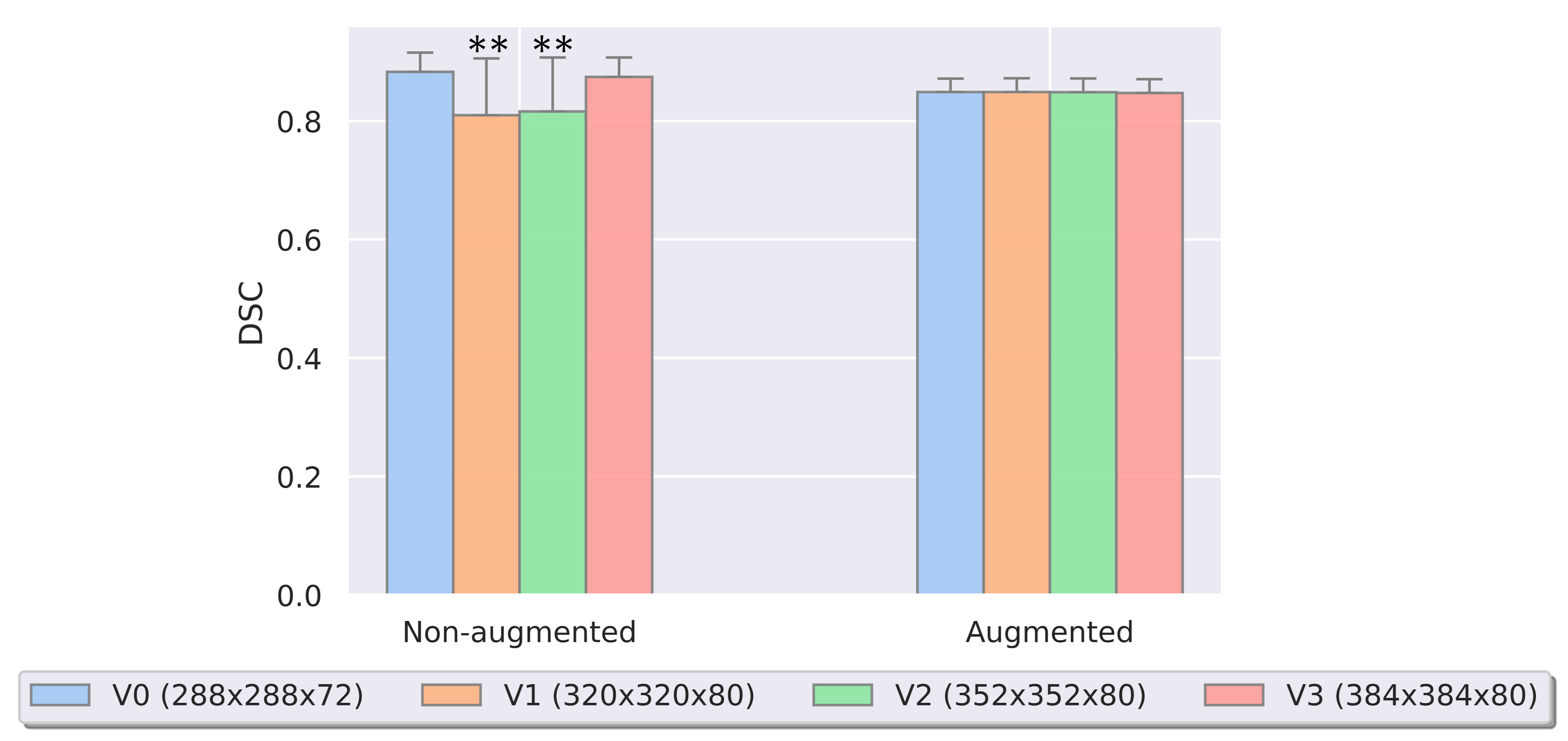}
\caption{The Dice score coefficient (DSC) accuracy on four test sets consisting of volumes at different spatial fields of view (FOVs) using the U-Nets trained on non-augmented and augmented training sets. Inference using the non-augmented U-Net is variable across different FOVs, with significantly lower accuracy in test sets cropped to different FOVs (p<0.01). While the augmented U-Net has a generally lower DSC on the test sets than the non-augmented U-Net, it performs consistently on volumes at different FOVs (p>0.99).}
\label{fig:fov-generalizability}
\end{figure}

\subsection{FOV Generalizability Comparison}
Baseline U-Net network performance was variable across test sets consisting of scans at different fields of view (Figure ~\ref{fig:fov-generalizability}). Inference on semi-cropped test sets (V1, V2) had significantly lower performance (p<0.01) than that on the original test set (V0). There was no significant difference (p=1.0) between performance on test set V0 and the non-cropped test set (V3). In contrast, there was no significant difference in performance of the augmented U-Net model across all four test sets (p>0.99).

\begin{figure}[bt]
\centering
\includegraphics[width=10cm]{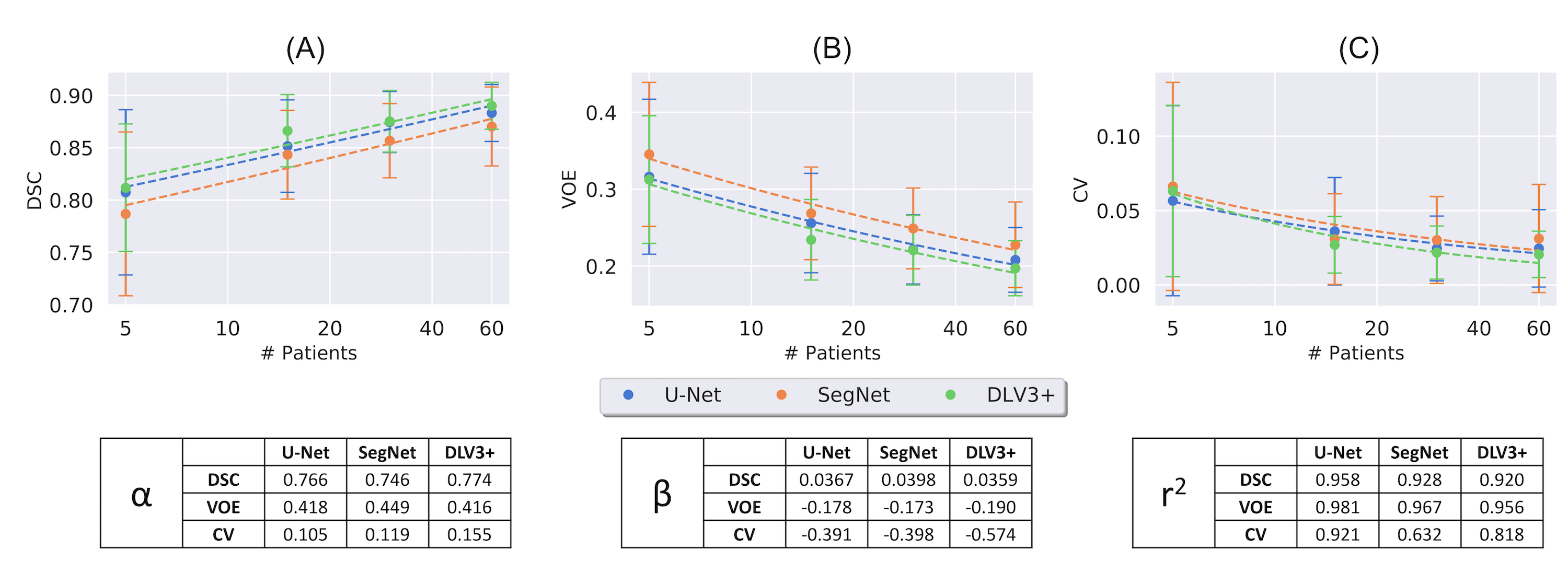}
\caption{Performance of U-Net, SegNet, and DeeplabV3+ (DLV3+) when trained on retrospectively subsampled training data. The plots (log-x scale) and corresponding $r^2$ values indicate a power-law relationship between segmentation performance, as measured by the (A) Dice score coefficient accuracy (DSC), (B) volumetric overlap error (VOE), and (C) coefficient of variation (CV), and the number of training patients for all networks. Experiments were repeated 3 times with fixed Python random seeds to ensure reproducibility of the results.}
\label{fig:data-limit}
\end{figure}

\subsection{Data Extent Trend}
Network performance for all three networks increased with increasing training data (Figure ~\ref{fig:data-limit}). The trend between the number of patients in the training dataset and network performance followed a power-law ($y=\alpha x^\beta$) scaling, as hypothesized previously \cite{amari1992four}, for all performance metrics (p<1e-4). Pixel-wise performance metrics, DSC and VOE, had a strong fit to the hypothesized power-law curve for all architectures ($r^2>0.91$ and $r^2>0.95$, respectively). CV had a relatively weaker, but still strong, fit ($r^2>0.63$). Among the different architectures, there was no significant difference in the intercept ($\alpha$) or exponent ($\beta$) of the curve fit measured at different seeds, and all exponents were less than 1 ($\beta$<1).

\section{Discussion}
In this study, we examined how variations in FCN architecture, loss functions, and training data impacted network performance for femoral cartilage segmentation. We found no significant pixel-wise difference in the performance of U-Net, SegNet, and DeeplabV3+, three commonly used FCN frameworks for natural image segmentation. There was also no significant performance difference between the segmentations produced by 2D and 2.5D networks. We demonstrated that BCE, soft Dice, and focal losses had similar false-positive and false-negative incidence rates, while WCE biased the network toward false-positive errors. Moreover, while data augmentation reduced U-Net performance, it increased generalizability in performance among scan volumes at different fields of view. Additionally, this study verified that segmentation performance scales directly, following a power-law relationship, with increasing data size. Traditionally, training methods and architectures have been a design choice when applying CNNs for semantic segmentation. In these cases, our findings provide insight into which design choices may be most effective for knee MR image segmentation using CNNs.

\subsection{Base Architecture Variations}
Based on network performance metrics, newer network architectures like DeeplabV3+ have slightly improved, though not significant, segmentation accuracy compared to traditionally used U-Net and SegNet models. The larger receptive field induced by using dilated convolutions in DeeplabV3+ may increase spatial awareness to foreground-background boundary regions. 

The \textit{expressivity} of a network, often used to characterize network generalizability, is defined as the degree to which the network structure facilitates learning features that are representative for the task. As expressivity increases, performance also increases. \textit{Raghu, et al.} and \textit{Bengio, et al.} suggest that expressivity is highly impacted by network structures such as depth, which enables hierarchical feature representations, and regularizations, which prime the network to learn representative features that are stable across different inputs \cite{raghu2016expressive, bengio2011expressive}. While DeeplabV3+ does not follow the same sequential autoencoder structure as U-Net and SegNet, it leverages dilated convolutions to extract features at various fields of view and decodes these features to create a hierarchical feature representation as expressive as those generated by the other two architectures.

Though network architecture has been closely linked with expressivity, there was no significant difference in the performance of the three network architectures, and all networks failed in similar regions of minimal, disjoint cartilage (Figure ~\ref{fig:arch-comparison}). The non-uniqueness in failure cases indicates that all three network models may optimize for similar deep features and consequently, segment images in a visually comparable manner. This minimal difference in performance suggests that beyond some threshold expressivity, differences in CNN architectures may have a negligible impact on the overall segmentation performance. Similar work for fully-connected neural networks (i.e. no convolutions) demonstrated network generalization is not limited by the architecture for a wide array of tasks, given that the network is expressive enough to achieve a small training error \cite{lampinen2018analytic}. While CNNs and fully-connected neural networks are not an exhaustive representation of all forms of neural networks, the trend of the limited effect of network structure on overall expressivity indicates that improving architectures may not be as effective in training better-performing networks.

\subsection{Practical Design for Volumetric Architectures}
In this study, the volumetric (2.5D/3D) networks had a negligible impact on segmentation accuracy and even performed worse than traditional 2D slice-wise segmentation in the case of the 3D network. The limited difference between 2.5D and 2D networks may be explained by the negligible difference in expressivity of these networks. These networks only differ at the first convolutional layer, which takes the image/volume as the input. While 2.5D networks accept an input volume ($M\times N\times t$), and 2D networks accept an input slice ($M\times N \times 1$), the output of the initial convolution layer is the same size in both networks. As a result, 2.5D networks only have more parameters in the first convolution layer (Supporting Figure S2), which is negligible when compared to the general size of the network and may not expressively represent the through-plane information 2.5D networks hope to capture.

Unlike 2.5D networks, which collapse the 3D input into multiple 2D features after the first convolutional layer, 3D networks maintain the depth-wise dimension throughout the network. While this allows depth-wise features to be extracted throughout the entire network, the number of network parameters also increases, which can limit the batch size of the network. In the 3D network trained in this study, a batch size of 1 was required to fit the scan volume as an input, which may have lead to less stable feature regularization. Additionally, to allow fitting the scan volume as an input, the 3D network had approximately the same number of parameters as the 2.5D and 3D networks. However, as the number of parameters per kernel increases to maintain the extra dimension, the number of filters at the initial convolutional layers had to be curbed twofold. The fewer filters at earlier stages in the network likely contributed to lower expressivity, and consequently poorer performance, of the network. With increased computational and parallelization power, designing 3D networks with similar filter counts as 2D networks may increase network expressivity.  

\subsection{Selecting Loss Functions}
While network architectures did not significantly impact performance, U-Net models trained using BCE, soft Dice, and focal losses performed significantly better than the model trained using WCE loss. While WCE is intended to normalize loss magnitude between imbalanced classes, the artificial weighting biases the network to over-classify the rarer class.

The degree of false-positive bias introduced into the network using WCE is likely modulated by the respective class weights. As the median frequency re-weighting method over-biases the network, traditional weighting protocols based on class incidence may not be the optimal weighting scheme. While optimal performance is traditionally measured by reducing the overall error, WCE loss weightings may be used to intentionally steer a network either towards additional false-positives or false-negatives, depending on the specific use case.

Additionally, the different error distributions around the threshold probability ($p_{FC}$=0.5) indicate the potential success of each loss function (Figure ~\ref{fig:loss-distribution}). In a binary problem, the probability output of pixel $i$ ($p_i$) is binarized at some threshold probability $p_T$, typically chosen to be the midpoint ($p_T$=0.5). Let $\hat{y}_i \in \{0,1\}$ define the output of the binarization operation $\beta$ on for pixel $i$, such that $
    \hat{y}_i = \beta(p_i) = \begin{cases}
        1, \;\; p_i>p_T\\
        0, \;\; p_i\leq p_T\\
\end{cases}$. Let $y_i \in \{0,1\}$ correspond to the ground-truth class for pixel $i$. Therefore, pixel $i$ is misclassified if $y_i \neq \beta(p_i)$. If pixel $i$ is misclassified, let $dp_i$ be the minimum amount of shift required to $p_i$ to correctly classify pixel $i$ (i.e. $\beta(p_i + dp_i)=y_i$). For the loss functions used above, the energy required to shift $p_i$ is directly proportional to $dp_i$. If $p_i$ is close to the limit bounds ($p_i \approx 0, 1$), $dp_i$ is very large; but if $p_i$ is \textit{close} to the threshold probability $p_T$, $dp_i$ is much smaller. Therefore, a distribution that is densely centered around $p_T$ minimizes $dp_i$ and has the most potential for reducing error rate with limited energy.

Of the four error distributions induced by different loss functions, focal loss produces errors that are most densely centered around $p_T$, which may make it most amenable for future optimization. Focal loss likely achieves this distribution by weighting the BCE loss to be inversely proportional to the correct classification. For example, a pixel with a probability for its correct class close to 1 will be weighted less than a pixel with a probability for its correct class close to 0. As a result, well-classified pixels will not contribute to the loss, and consequently, will not be further optimized. This preserves high error rate close to $p_{FC}$=0.5, as a network trained with focal loss is most uncertain about these examples. This symmetric distribution also suggests that correcting false-positive and false-negative errors would require an equal amount of energy.

\subsection{Achieving Multi-FOV Robustness through Data Augmentation}
As MR scan protocols can often adjust the image FOV for different sized patients, training an FCN that is generalizable to multiple FOVs may be desired. The U-Net trained on non-augmented images did not exhibit the same performance across different FOVs. Recall that test sets V1, V2, and V3 covered a larger through-plane field of view (80 slices) than V0, whose dimensions were identical to the training volumes (72 slices). Failure cases in V1 and V2 were predominately in the 8 slices not included in the volumetrically cropped testing/training sets. It is likely the network failed in these regions because these additional slices include anatomy that may not have been seen during training.

In contrast, the U-Net trained using augmented training data exhibited the same performance across all FOVs. The augmentations introduced realistic variations that could occur during imaging (motion and gamma variations) and artificial variations that change the distribution of anatomy across pixels (zooming and shearing). The later set of artificial augmentations manipulate the FOV that the tissue of interest covers in the training image. As a result, the optimized network likely consists of a family of features that is robust to spatial FOV variations within the degrees of the zooming and shearing distributions used. Thus, instead of measuring the expressivity of a FCN network on a single test set, we suggest that the expressivity for multi-FOV applications should be quantified by its performance on test sets at varying FOVs for evaluating robustness to multi-FOV scans.

While augmentations have been readily accepted as a method to increase network accuracy, the 2D U-Net trained with augmented data in this study had sub-optimal performance. This phenomenon likely occurred because the network trained with non-augmented data optimized features for images containing the same FOV of anatomy as the training images. In contrast, the augmented dataset may challenge the network by varying the FOV and contrast of information seen. The optimal minimum may not minimize loss as efficiently on the non-augmented datasets, and as a result, the features are not optimized to achieve a high testing loss on test set V0. However, these features likely increased the stability of the network for inference on scans of multiple FOVs. We suggest that augmentations should be meticulously curated to increase network expressivity to expected image variations, especially in regards to tissues of interest having variable sizes in potential test images.

\subsection{Navigating Training with Limited Data}
The performance of all three networks changed at a considerably slow rate as data size increased. The rate is primarily governed by the exponent value ($\beta$) in the power-law equation. The mean exponent across three seeds $\bar{\beta}$<0.05 for all architectures indicated a slow growth in performance as a function of data size. In a recent work, Hestness, et al. empirically verified that the error rate in image classification decreases following a power-law scaling with $\beta<1$ regardless of architecture \cite{hestness2017deep}. Like image classification, semantic segmentation also appeared to follow this trend, with minimal variation in $\beta$ among architectures.

Moreover, this slow-growth power-law performance scaling can allow us to empirically estimate the performance of these networks as the data size increases. Based on these parameter estimates, achieving a 95\% Dice accuracy for the U-Net, SegNet, and DeeplabV3+ models would require approximately 350, 440, and 300 patients, respectively. Therefore, while increasing training data does increase performance over time, the addition of each subsequent dataset diminishes marginal utility. These results suggest that even with small amounts of data, high percentage of performance can generally be obtained.

\subsection{Limitations}
Despite the promising empirical relationships elucidated in this work, there were limitations to this study that should be addressed in future studies. Training hyperparameters for each network were empirically determined by investigating training loss curves for the initial epochs. While a robust hyperparameter search may yield a more optimal set for training, this was beyond the primary premise of this work, which aimed to explore larger tradeoffs between network architectures, loss functions, and training data. Additionally, the 3D U-Net architecture trained in the volumetric architecture experiment fixed the input depth at 32 slices, resulting in a low batch size and fewer number of filters at each network level. Future studies could modulate the number of input slices to increase batch size and number of filters to optimize network performance. Moreover, all networks performed binary segmentation, but as most loss functions allow for multi-class segmentation, it would be useful to understand the impact of this problem on performance for each tissue.

\section{Conclusion}
In this study, we quantified the impact of variations in network architecture, loss functions, and training data for segmenting femoral cartilage from 3D MRI in order to investigate the tradeoffs involved in segmentation with CNNs. Variations in network architectures yielded minimal differences in overall segmentation accuracy. Additionally, loss functions dictate how the network weights are optimized and, as a result, influence how errors are distributed across probabilities. Moreover, realistic data augmentation methods can increase network generalizability at the cost of absolute network performance on any given test set. Limited amounts of training data may also not be the bottleneck in network performance.

\subsubsection*{Acknowledgments}
Contract grant sponsor: National Institutes of Health (NIH); contract grant numbers NIH R01 AR063643, R01 EB002524, K24 AR062068, and P41 EB015891. Contract grant sponsor: Philips (research support). Image data was acquired from the Osteoarthritis Initiative (OAI). The OAI is a public-private partnership comprised of five contracts (N01-AR-2-2258; N01-AR-2-2259; N01-AR-2-2260; N01-AR-2-2261; N01-AR-2-2262) funded by the National Institutes of Health, a branch of the Department of Health and Human Services, and conducted by the OAI Study Investigators. Private funding partners include Merck Research Laboratories; Novartis Pharmaceuticals Corporation, GlaxoSmithKline; and Pfizer, Inc. Private sector funding for the OAI is managed by the Foundation for the National Institutes of Health. This manuscript was prepared using an OAI public use data set and does not necessarily reflect the opinions or views of the OAI investigators, the NIH, or the private funding partners.

\bibliography{ms}

\begin{thebibliography}{10}

\bibitem{hoyte2011segmentations}
Hoyte~L, Ye~W, Brubaker~L, Fielding~JR, Lockhart~ME, Heilbrun~ME, Brown~MB,
  Warfield~SK, Network~PFD.
\newblock Segmentations of mri images of the female pelvic floor: A study of
  inter-and intra-reader reliability.
\newblock Journal of Magnetic Resonance Imaging 2011; 33:684--691.

\bibitem{bogner2012readout}
Bogner~W, PinkerDomenig~K, Bickel~H, Chmelik~M, Weber~M, Helbich~TH,
  Trattnig~S, Gruber~S.
\newblock Readout-segmented echo-planar imaging improves the diagnostic
  performance of diffusion-weighted mr breast examinations at 3.0 t.
\newblock Radiology 2012; 263:64--76.

\bibitem{dam2015automatic}
Dam~EB, Lillholm~M, Marques~J, Nielsen~M.
\newblock Automatic segmentation of high-and low-field knee mris using knee
  image quantification with data from the osteoarthritis initiative.
\newblock Journal of Medical imaging 2015; 2:024001.

\bibitem{bauer2013survey}
Bauer~S, Wiest~R, Nolte~LP, Reyes~M.
\newblock A survey of mri-based medical image analysis for brain tumor studies.
\newblock Physics in Medicine \& Biology 2013; 58:R97.

\bibitem{erhart2015new}
ErhartHledik~JC, Favre~J, Andriacchi~TP.
\newblock New insight in the relationship between regional patterns of knee
  cartilage thickness, osteoarthritis disease severity, and gait mechanics.
\newblock Journal of biomechanics 2015; 48:3868--3875.

\bibitem{dunn2004t2}
Dunn~TC, Lu~Y, Jin~H, Ries~MD, Majumdar~S.
\newblock T2 relaxation time of cartilage at mr imaging: comparison with
  severity of knee osteoarthritis.
\newblock Radiology 2004; 232:592--598.

\bibitem{eckstein2013quantitative}
Eckstein~F, Kwoh~CK, Boudreau~RM, Wang~Z, Hannon~MJ, Cotofana~S, Hudelmaier~MI,
  Wirth~W, Guermazi~A, Nevitt~MC et~al.
\newblock Quantitative mri measures of cartilage predict knee replacement: a
  case--control study from the osteoarthritis initiative.
\newblock Annals of the Rheumatic Diseases 2013; 72:707--714.

\bibitem{welsch2008cartilage}
Welsch~GH, Mamisch~TC, Domayer~SE, Dorotka~R, KutschaLissberg~F, Marlovits~S,
  White~LM, Trattnig~S.
\newblock Cartilage t2 assessment at 3-t mr imaging: in vivo differentiation of
  normal hyaline cartilage from reparative tissue after two cartilage repair
  procedures—initial experience.
\newblock Radiology 2008; 247:154--161.

\bibitem{Eckstein_2006}
Eckstein~F.
\newblock Double echo steady state magnetic resonance imaging of knee articular
  cartilage at 3 tesla: a pilot study for the osteoarthritis initiative.
\newblock Annals of the Rheumatic Diseases 2006; 65:433--441.

\bibitem{seim2010model}
Seim~H, Kainmueller~D, Lamecker~H, Bindernagel~M, Malinowski~J, Zachow~S.
\newblock Model-based auto-segmentation of knee bones and cartilage in mri
  data.
\newblock Medical Image Analysis for the Clinic: A Grand Challenge 2010; pp.
  215--223.

\bibitem{shan2014automatic}
Shan~L, Zach~C, Charles~C, Niethammer~M.
\newblock Automatic atlas-based three-label cartilage segmentation from mr knee
  images.
\newblock Medical image analysis 2014; 18:1233--1246.

\bibitem{pedoia2016segmentation}
Pedoia~V, Majumdar~S, Link~TM.
\newblock Segmentation of joint and musculoskeletal tissue in the study of
  arthritis.
\newblock Magnetic Resonance Materials in Physics, Biology and Medicine 2016;
  29:207--221.

\bibitem{liu2018deep}
Liu~F, Zhou~Z, Jang~H, Samsonov~A, Zhao~G, Kijowski~R.
\newblock Deep convolutional neural network and 3d deformable approach for
  tissue segmentation in musculoskeletal magnetic resonance imaging.
\newblock Magnetic resonance in medicine 2018; 79:2379--2391.

\bibitem{norman2018use}
Norman~B, Pedoia~V, Majumdar~S.
\newblock Use of 2d u-net convolutional neural networks for automated cartilage
  and meniscus segmentation of knee mr imaging data to determine relaxometry
  and morphometry.
\newblock Radiology 2018; p. 172322.

\bibitem{lecun2015deep}
LeCun~Y, Bengio~Y, Hinton~G.
\newblock Deep learning.
\newblock nature 2015; 521:436.

\bibitem{shwartz2017opening}
ShwartzZiv~R, Tishby~N.
\newblock Opening the black box of deep neural networks via information.
\newblock arXiv preprint arXiv:1703.00810 2017; .

\bibitem{russakovsky2015imagenet}
Russakovsky~O, Deng~J, Su~H, Krause~J, Satheesh~S, Ma~S, Huang~Z, Karpathy~A,
  Khosla~A, Bernstein~M et~al.
\newblock Imagenet large scale visual recognition challenge.
\newblock International Journal of Computer Vision 2015; 115:211--252.

\bibitem{szegedy2015going}
Szegedy~C, Liu~W, Jia~Y, Sermanet~P, Reed~S, Anguelov~D, Erhan~D, Vanhoucke~V,
  Rabinovich~A.
\newblock Going deeper with convolutions.
\newblock { In:} Proceedings of the IEEE conference on computer vision and
  pattern recognition, { In:} Proceedings of the IEEE conference on computer
  vision and pattern recognition, 2015.  pp. 1--9.

\bibitem{he2016deep}
He~K, Zhang~X, Ren~S, Sun~J.
\newblock Deep residual learning for image recognition.
\newblock { In:} Proceedings of the IEEE conference on computer vision and
  pattern recognition, { In:} Proceedings of the IEEE conference on computer
  vision and pattern recognition, 2016.  pp. 770--778.

\bibitem{kamnitsas2017efficient}
Kamnitsas~K, Ledig~C, Newcombe~VF, Simpson~JP, Kane~AD, Menon~DK, Rueckert~D,
  Glocker~B.
\newblock Efficient multi-scale 3d cnn with fully connected crf for accurate
  brain lesion segmentation.
\newblock Medical image analysis 2017; 36:61--78.

\bibitem{cciccek20163d}
{\c{C}}i{\c{c}}ek~{\"O}, Abdulkadir~A, Lienkamp~SS, Brox~T, Ronneberger~O.
\newblock 3d u-net: learning dense volumetric segmentation from sparse
  annotation.
\newblock { In:} International Conference on Medical Image Computing and
  Computer-Assisted Intervention, { In:} International Conference on Medical
  Image Computing and Computer-Assisted Intervention, 2016.  pp. 424--432.

\bibitem{tian2017deep}
Tian~Z, Liu~L, Fei~B.
\newblock Deep convolutional neural network for prostate mr segmentation.
\newblock { In:} Medical Imaging 2017: Image-Guided Procedures, Robotic
  Interventions, and Modeling, { In:} Medical Imaging 2017: Image-Guided
  Procedures, Robotic Interventions, and Modeling, 2017.  p. 101351L.

\bibitem{baumgartner2017exploration}
Baumgartner~CF, Koch~LM, Pollefeys~M, Konukoglu~E.
\newblock An exploration of 2d and 3d deep learning techniques for cardiac mr
  image segmentation.
\newblock { In:} International Workshop on Statistical Atlases and
  Computational Models of the Heart, { In:} International Workshop on
  Statistical Atlases and Computational Models of the Heart, 2017.  pp.
  111--119.

\bibitem{perez2017effectiveness}
Perez~L, Wang~J.
\newblock The effectiveness of data augmentation in image classification using
  deep learning.
\newblock arXiv preprint arXiv:1712.04621 2017; .

\bibitem{krizhevsky2012imagenet}
Krizhevsky~A, Sutskever~I, Hinton~GE.
\newblock Imagenet classification with deep convolutional neural networks.
\newblock { In:} Advances in neural information processing systems, { In:}
  Advances in neural information processing systems, 2012.  pp. 1097--1105.

\bibitem{long2015fully}
Long~J, Shelhamer~E, Darrell~T.
\newblock Fully convolutional networks for semantic segmentation.
\newblock { In:} Proceedings of the IEEE conference on computer vision and
  pattern recognition, { In:} Proceedings of the IEEE conference on computer
  vision and pattern recognition, 2015.  pp. 3431--3440.

\bibitem{Peterfy_2008}
Peterfy~C, Schneider~E, Nevitt~M.
\newblock The osteoarthritis initiative: report on the design rationale for the
  magnetic resonance imaging protocol for the knee.
\newblock Osteoarthritis and Cartilage 2008; 16:1433--1441.

\bibitem{kellgren1958osteo}
Kellgren~J, Lawrence~J.
\newblock Osteo-arthrosis and disk degeneration in an urban population.
\newblock Annals of the Rheumatic Diseases 1958; 17:388.

\bibitem{ronneberger2015u}
Ronneberger~O, Fischer~P, Brox~T.
\newblock U-net: Convolutional networks for biomedical image segmentation.
\newblock { In:} International Conference on Medical image computing and
  computer-assisted intervention, { In:} International Conference on Medical
  image computing and computer-assisted intervention, 2015.  pp. 234--241.

\bibitem{badrinarayanan2015segnet}
Badrinarayanan~V, Kendall~A, Cipolla~R.
\newblock Segnet: A deep convolutional encoder-decoder architecture for image
  segmentation.
\newblock arXiv preprint arXiv:1511.00561 2015; .

\bibitem{chen2018deeplab}
Chen~LC, Papandreou~G, Kokkinos~I, Murphy~K, Yuille~AL.
\newblock Deeplab: Semantic image segmentation with deep convolutional nets,
  atrous convolution, and fully connected crfs.
\newblock IEEE transactions on pattern analysis and machine intelligence 2018;
  40:834--848.

\bibitem{glorot2010understanding}
Glorot~X, Bengio~Y.
\newblock Understanding the difficulty of training deep feedforward neural
  networks.
\newblock { In:} Proceedings of the thirteenth international conference on
  artificial intelligence and statistics, { In:} Proceedings of the thirteenth
  international conference on artificial intelligence and statistics, 2010.
  pp. 249--256.

\bibitem{pascanu2013difficulty}
Pascanu~R, Mikolov~T, Bengio~Y.
\newblock On the difficulty of training recurrent neural networks.
\newblock { In:} International Conference on Machine Learning, { In:}
  International Conference on Machine Learning, 2013.  pp. 1310--1318.

\bibitem{chollet2017xception}
Chollet~F.
\newblock Xception: Deep learning with depthwise separable convolutions.
\newblock arXiv preprint 2017; pp. 1610--02357.

\bibitem{chen2017rethinking}
Chen~LC, Papandreou~G, Schroff~F, Adam~H.
\newblock Rethinking atrous convolution for semantic image segmentation.
\newblock arXiv preprint arXiv:1706.05587 2017; .

\bibitem{roth2014new}
Roth~HR, Lu~L, Seff~A, Cherry~KM, Hoffman~J, Wang~S, Liu~J, Turkbey~E,
  Summers~RM.
\newblock A new 2.5 d representation for lymph node detection using random sets
  of deep convolutional neural network observations.
\newblock { In:} International conference on medical image computing and
  computer-assisted intervention, { In:} International conference on medical
  image computing and computer-assisted intervention, 2014.  pp. 520--527.

\bibitem{milletari2016v}
Milletari~F, Navab~N, Ahmadi~SA.
\newblock V-net: Fully convolutional neural networks for volumetric medical
  image segmentation.
\newblock { In:} 3D Vision (3DV), 2016 Fourth International Conference on, {
  In:} 3D Vision (3DV), 2016 Fourth International Conference on, 2016.  pp.
  565--571.

\bibitem{lin2018focal}
Lin~TY, Goyal~P, Girshick~R, He~K, Doll{\'a}r~P.
\newblock Focal loss for dense object detection.
\newblock IEEE transactions on pattern analysis and machine intelligence 2018;
  .

\bibitem{rafiq2001neural}
Rafiq~M, Bugmann~G, Easterbrook~D.
\newblock Neural network design for engineering applications.
\newblock Computers \& Structures 2001; 79:1541--1552.

\bibitem{nair2010rectified}
Nair~V, Hinton~GE.
\newblock Rectified linear units improve restricted boltzmann machines.
\newblock { In:} Proceedings of the 27th international conference on machine
  learning (ICML-10), { In:} Proceedings of the 27th international conference
  on machine learning (ICML-10), 2010.  pp. 807--814.

\bibitem{he2015delving}
He~K, Zhang~X, Ren~S, Sun~J.
\newblock Delving deep into rectifiers: Surpassing human-level performance on
  imagenet classification.
\newblock { In:} Proceedings of the IEEE international conference on computer
  vision, { In:} Proceedings of the IEEE international conference on computer
  vision, 2015.  pp. 1026--1034.

\bibitem{kingma2014adam}
Kingma~DP, Ba~J.
\newblock Adam: A method for stochastic optimization.
\newblock arXiv preprint arXiv:1412.6980 2014; .

\bibitem{abadi2016tensorflow}
Abadi~M, Barham~P, Chen~J, Chen~Z, Davis~A, Dean~J, Devin~M, Ghemawat~S,
  Irving~G, Isard~M et~al.
\newblock Tensorflow: a system for large-scale machine learning.
\newblock { In:} OSDI, { In:} OSDI, 2016.  pp. 265--283.

\bibitem{taha2015metrics}
Taha~AA, Hanbury~A.
\newblock Metrics for evaluating 3d medical image segmentation: analysis,
  selection, and tool.
\newblock BMC medical imaging 2015; 15:29.

\bibitem{jones2014scipy}
Jones~E, Oliphant~T, Peterson~P.
\newblock $\{$SciPy$\}$: open source scientific tools for $\{$Python$\}$.
\newblock 2014; .

\bibitem{amari1992four}
Amari~Si, Fujita~N, Shinomoto~S.
\newblock Four types of learning curves.
\newblock Neural Computation 1992; 4:605--618.

\bibitem{raghu2016expressive}
Raghu~M, Poole~B, Kleinberg~J, Ganguli~S, SohlDickstein~J.
\newblock On the expressive power of deep neural networks.
\newblock arXiv preprint arXiv:1606.05336 2016; .

\bibitem{bengio2011expressive}
Bengio~Y, Delalleau~O.
\newblock On the expressive power of deep architectures.
\newblock { In:} International Conference on Algorithmic Learning Theory, {
  In:} International Conference on Algorithmic Learning Theory, 2011.  pp.
  18--36.

\bibitem{lampinen2018analytic}
Lampinen~AK, Ganguli~S.
\newblock An analytic theory of generalization dynamics and transfer learning
  in deep linear networks.
\newblock arXiv preprint arXiv:1809.10374 2018; .

\bibitem{hestness2017deep}
Hestness~J, Narang~S, Ardalani~N, Diamos~G, Jun~H, Kianinejad~H, Patwary~M,
  Ali~M, Yang~Y, Zhou~Y.
\newblock Deep learning scaling is predictable, empirically.
\newblock arXiv preprint arXiv:1712.00409 2017; .

\end{thebibliography}
\bibliographystyle{mrm}

\end{document}


\maketitle

\section{2.5D Networks}
A 2.5D network takes a 3D volume of thickness $t$ as an input and produces a 2D segmentation mask for the central slice. For example, if $t=5$, the stack of slices \{$1$, $2$, $3$, $4$, $5$\} would be the 3D input into the network to produce a 2D segmentation mask for slice $3$. In the case of edge slices, slices extending beyond the volume were repeated. Such a 2.5D approach can provide the CNN through-plane contextual features, while avoiding the use of computationally complex 3D convolutions. 

Let $S_i$ be the slice in a volume of $|V|$ slices indexed by $i$ such that $i=\{1, 2, 3, ..., |V|\}$. Let $I(s, t)$ be a 3D input of thickness $t$ to produce a 2D segmentation mask for slice $s$ as seen in Eq. \ref{eq:2.5D}:

\begin{equation}
\label{eq:2.5D}
    I(s, t) = \{S_{max(s+j, 1)} | j=-t // 2, ..., -1\} \cup \{S_s\} \cup \{S_{min(s+k, |V|)} | k=\{1, ..., t // 2\}
\end{equation}

The $//$ operation refers to integer division. Input thicknesses of $t=3, 5, 7$ were used.

\section{Pixel-wise Loss Functions}
Pixel-wise loss functions aggregate the penalty of misclassifying each pixel in the target segmentation mask. In this experiment, we utilized four pixel-wise losses: binary cross-entropy (BCE) (Eq. \ref{eq:bce}), soft dice loss (Eq. \ref{eq:dice_loss}), weighted cross entropy (WCE) (Eq. \ref{eq:wce}), and focal loss ($\gamma$=3) (Eq. \ref{eq:focal_loss}).

\begin{equation}
\label{eq:bce}
BCE = \frac{1}{N}\sum_{n=1}^{N}\sum_{c=0}^{1} y_{n, c} log(p_{n, c})
\end{equation}

\begin{equation}
\label{eq:dice_loss}
Dice = \frac{\sum_{n=1}^{N} 2 y_{n} p_{n}}{\sum_{n=1}^{N} (y_n + p_n)}
\end{equation}

\begin{equation}
\label{eq:wce}
WCE = \frac{1}{N}\sum_{n=1}^{N}\sum_{c=0}^{1} w_c y_{n, c} log(p_{n, c})
\end{equation}

\begin{equation}
\label{eq:focal_loss}
Focal = \frac{1}{N}\sum_{n=1}^{N} -(1-p_t)^\gamma log(p_t) \quad
p_t = \begin{cases}
        p_{n,1}, \;\; y_{n,1}=1 \;(y_{n,0}=0)\\
        1-p_{n,1}, \;\; y_{n,1}=0 \; (y_{n,0}=1)\\
\end{cases}
\end{equation}

In Eqs. \ref{eq:bce}-\ref{eq:focal_loss}, $n$ indexes the image elements (pixels) and $c$ indexes the two classes (0- background, 1- femoral cartilage). For WCE (Eq. \ref{eq:wce}), the class weights ($w_c$) were defined by median frequency re-weighting, $w_c = \frac{median(f_0, f_1)}{f_c}$ where $f_c$ refers to the frequency of pixels corresponding to class $c$ in the training set. The normalized weights were empirically found to be $w_0=1.0$ and $w_1=75.6$. $y_{n}$ is a one-hot-encoded vector corresponding to the ground truth class of pixel $n$. $p_{n, c}$ refers to the probability that pixel $n$ is of class $c$. Note that because this is a binary problem, $p_{n,0} + p_{n,1} = 1$.

\clearpage
\section*{Supplemental Graphics}
\begin{table}[H]
\renewcommand\thetable{S1} 
\centering
\includegraphics[width=7cm]{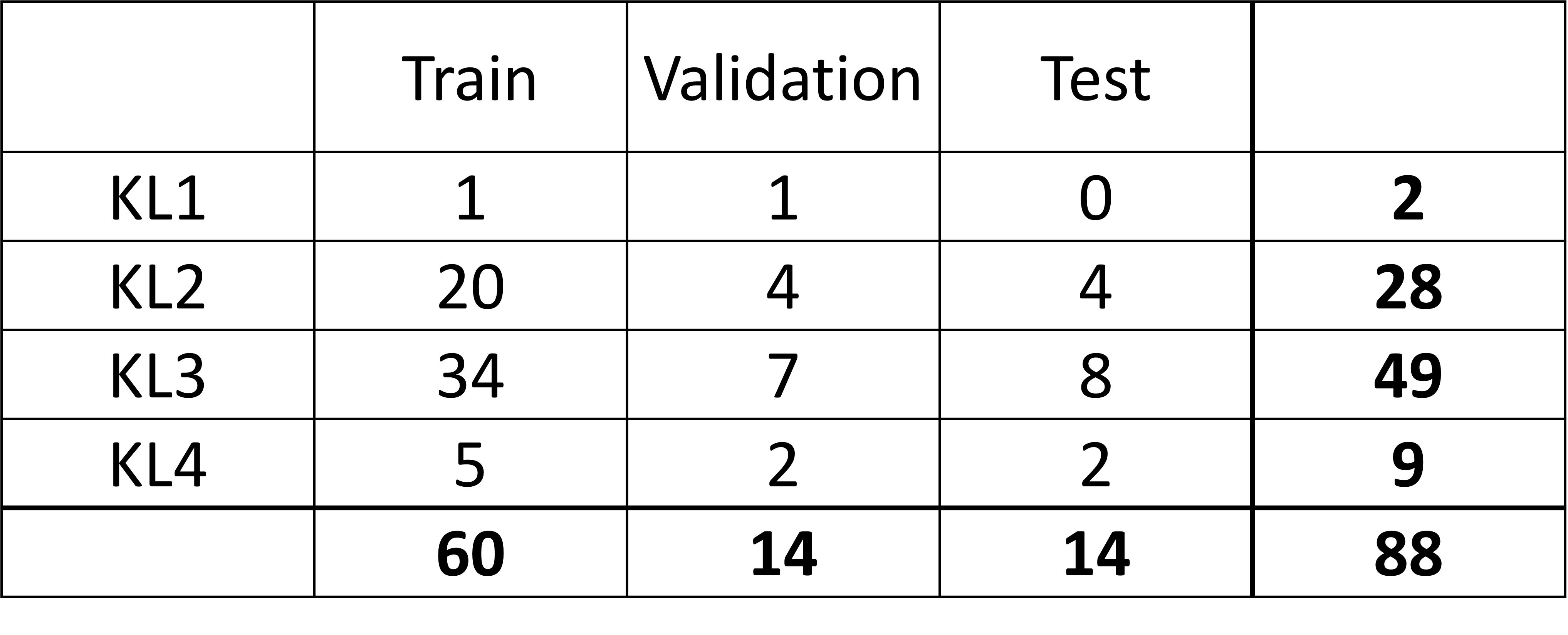}
\caption{Distribution of patients by Kellgren-Lawrence (KL) grade at baseline scan in training, validation, and testing sets.}
\label{supp-fig:KL-set-distribution}
\end{table}

\begin{figure}[H]
\renewcommand\thefigure{S2} 
\centering
\includegraphics[width=10cm]{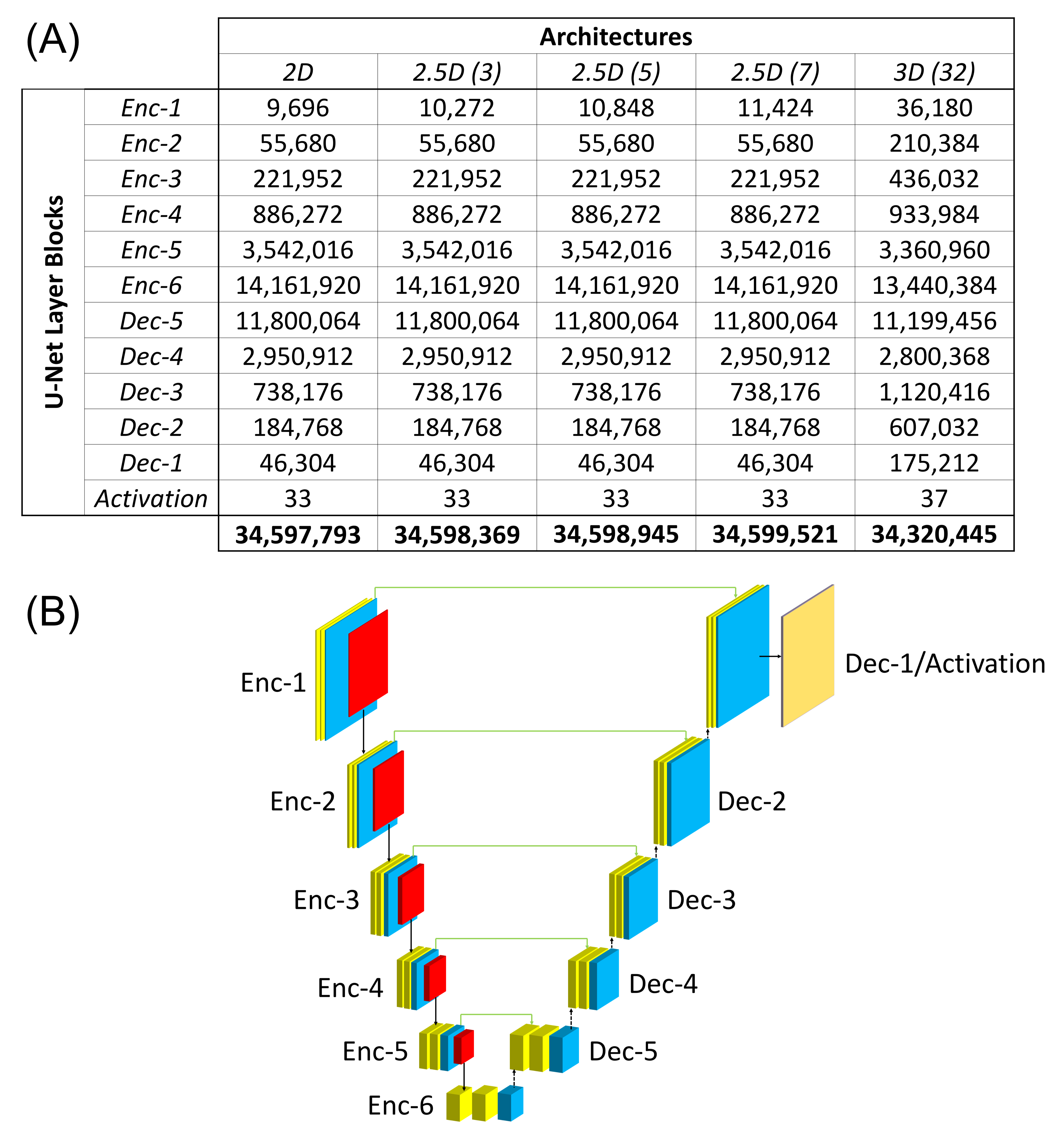}
\caption{(A) Number of weights for volumetric U-Net models (2D/2.5D/3D) at different encoder (enc) and decoder (dec) depths. Encoder/decoder layer blocks at different depths are indicated in (B). Input thickness (i.e. number of input slices) for 2.5D and 3D networks are specified in parentheses.}
\label{supp-fig:volume-arch-params}
\end{figure}

\begin{table}[H]
\renewcommand\thetable{S3} 
\centering
\includegraphics[width=7cm]{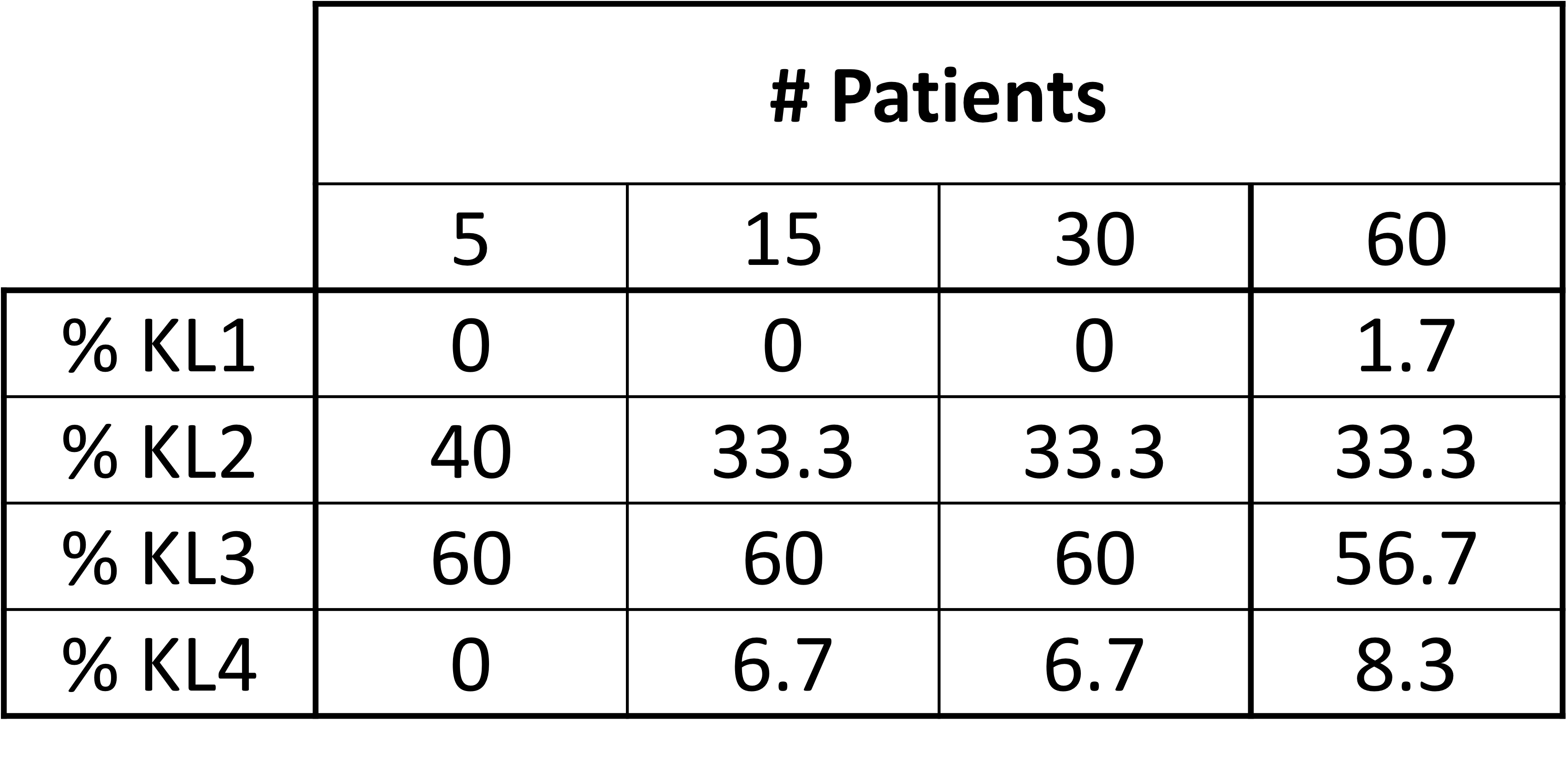}
\caption{Distribution of patients by Kellgren-Lawrence (KL) grades at baseline scan in retrospectively subsampled training sets. KL distributions were closely maintained across subsampled training sets.}
\label{supp-fig:data-limit-distribution}
\end{table}